\documentclass[fleqn,usenatbib]{mnras}

\usepackage{newtxtext,newtxmath}

\usepackage[T1]{fontenc}
\usepackage{float}

\DeclareRobustCommand{\VAN}[3]{#2}
\let\VANthebibliography\thebibliography
\def\thebibliography{\DeclareRobustCommand{\VAN}[3]{##3}\VANthebibliography}

\usepackage{graphicx}	
\usepackage{amsmath}	
\usepackage{relsize}

\graphicspath{ {./images/} }
\usepackage{bm}
\usepackage{multirow} 
\usepackage{gensymb} 
\usepackage{placeins} 

\defcitealias{EHT1}{EHT~I}
\defcitealias{EHT2}{EHT~II}
\defcitealias{EHT3}{EHT~III}
\defcitealias{EHT4}{EHT~IV}
\defcitealias{EHT5}{EHT~V}
\defcitealias{EHT6}{EHT~VI}

\defcitealias{Lockhart-Gralla-2021}{Paper~I}

\title[M87* Modeling]{
How narrow is the M87* ring? II. A new geometric model
}

\author[
Lockhart and Gralla
]{
Will Lockhart$^{1}$\thanks{wlockhart@email.arizona.edu}
and Samuel E. Gralla,$^{1}$
\\
$^{1}$Department of Physics, University of Arizona, Tucson AZ, USA
}

\begin{document}
\label{firstpage}
\pagerange{\pageref{firstpage}--\pageref{lastpage}}
\maketitle

\begin{abstract}

The 2017 Event Horizon Telescope (EHT) observations of M87* detected a ring-shaped feature $\sim40\mu$as in diameter, consistent with the event horizon scale of a black hole of the expected mass. The thickness of this ring, however, proved difficult to measure, despite being an important parameter for constraining the observational appearance. In the first paper of this series we asked whether the width of the ring was sensitive to the choice of likelihood function used to compare observed closure phases and closure amplitudes to model predictions. In this paper we investigate whether the ring width is robust to changes in the model itself. We construct a more realistic geometric model with two new features: an adjustable radial falloff in brightness, and a secondary ``photon ring'' component in addition to the primary annulus. This thin, secondary ring is predicted by gravitational lensing for any black hole with an optically thin accretion flow. Analyzing the data using the new model, we find that the primary annulus remains narrow (fractional width $\leq0.25$) even with the added model freedom. This provides further evidence in favor of a narrow ring for the true sky appearance of M87*, a surprising feature that, if confirmed, would demand theoretical explanation. Comparing the Bayesian evidence for models with and without a secondary ring, we find no evidence for the presence of a lensed photon ring in the 2017 observations.  However, the techniques we introduce may prove useful for future observations with a larger and more sensitive array. 
\vspace{4mm}
\end{abstract}

\begin{keywords}
galaxies: nuclei -- (galaxies:) quasars: supermassive black holes -- submillimetre: galaxies -- black hole physics -- techniques: interferometric
\vspace{-3mm}
\end{keywords}


\section{Introduction}

The initial analysis of the 2017 Event Horizon Telescope (EHT) data demonstrated that the observational appearance of M87* is predominantly ring-shaped \citep[henceforth EHT I-VI]{EHT1,EHT2,EHT3,EHT4,EHT5,EHT6}. The gross features of the image -- the presence of a central brightness depression, the diameter of the surrounding annulus (approximately 40 micro-arcseconds ($\mu$as) across), as well as the direction of a brightness asymmetry -- were all found to be robust across different analysis methods and observation days. These features agree with models of synchrotron emission from matter accreting onto a black hole, and provide the most direct evidence to date that M87* is indeed a supermassive black hole.

In contrast, the analysis was unable to constrain the \textit{thickness} of the annulus to the same degree.  This parameter is important because the radial profile of the image is directly related to the emission profile of the accretion disk.  Although light emitted near the horizon is bent significantly by the gravitational field on its way to the observer, rays from the from the foreground / mid-plane region remain mostly parallel, so that the actual distortion of the image is small (see \citetalias{Lockhart-Gralla-2021}, Sec 2.2). Since the radial brightness pattern arrives relatively intact, constraining it could ultimately help to discriminate between different models of the accretion flow around M87*. 

In  \citetalias{Lockhart-Gralla-2021} of this series we pointed out that the fractional width of $f_w \sim 0.25$ inferred from geometric modeling is on the lower end of expectation and differs significantly from the widths inferred from the EHT imaging pipeline, while at least some modeling results were so thin that they are inconsistent with predictions. In that paper we re-examined the data using a slightly different likelihood function in the analysis and found that, while the diameter and brightness gradient were unaffected, the width of the ring \textit{was} sensitive to this change, underscoring the difficulty of pinning down this parameter.  

Past modeling efforts by the authors \citepalias{Lockhart-Gralla-2021} and by the EHT collaboration \citepalias{EHT6} used a geometric model called ``\textsf{xs-ring}''.\footnote{A nearly identical model \textsf{xs-ringauss} was also explored in \citetalias{EHT6}.} In our view, while \textsf{xs-ring} is well-suited for determining whether the image must be ring-like (i.e., must have a central brightness depression), it is not necessarily adequate for constraining the radial profile.  We will highlight two drawbacks of this model: a necessarily fast brightness falloff, and the lack of a thin, `photon ring' component.  Does the strictly prescribed brightness falloff bias the results? Might it be that the \textsf{xsring} model produces thin rings because it is trying to simultaneously fit a wider annulus and a much narrower photon ring?  To address these questions, here we design a new geometric model in an effort to better understand the constraints on the width of the M87* annulus.

The \textsf{xs-ring} model has two major limitations. The first is that the radial falloff in brightness on the outer edge of the ring is always extremely fast. The shape of the outer edge is the result of blurring a step function with a Gaussian kernel, so the decay is necessarily Gaussian. The emissivity of the accretion disk, however, may decrease with distance much more slowly than that. A recent study concluded that the decay is likely to be exponential under the conditions expected near M87* \citep{Vincent2022}, and power-law decay is also a reasonable possibility. It is therefore quite natural to ask whether allowing for slower radial falloff in the model is necessary to properly constrain the width of the ring.  

The second limitation of \textsf{xs-ring} is that it does not include a photon ring.  If the accreting material around M87* is optically thin, as is widely believed \citepalias{EHT1}, then the image must contain a lensed component coming from photons that orbit the black hole (possibly multiple times) before reaching the observer \citep{luminet1979}. These trajectories all appear on the image plane near a critical curve, where photons sent backwards from the image plane would asymptote to a bound orbit \citep{bardeen1973}. For a disk-like source, this lensed component takes the shape of a narrow ring known as the \textit{photon ring}.\footnote{In general there will be a sequence of increasingly narrow photon rings converging to the critical curve.  The higher order rings contain negligible flux density \citep{gralla-holz-wald2019} and are ignored in this study.} Because the critical curve is a property of the black hole spacetime only, independent of the nature of the surrounding matter, the diameter of the photon ring is a robust measure of the mass-over-distance ratio of the black hole.  Detecting the photon ring would verify a basic prediction of general relativity, while measuring its diameter would provide a black hole mass measurement independent of astrophysical assumptions. 

The width of the photon ring in most models is only $\sim 2$--$3\mu$as [e.g., \citet[Fig.~3]{johnson-etal2020}, or \citet[ Fig.~7]{chael-johnson-lupsasca2021}], which is well below the nominal EHT imaging resolution of $\sim 20\mu$as. However, it may still be possible to infer its presence from fitting models in the visibility domain.  A useful case study is the lensed dusty star-forming galaxy SPT0346-52, in which lens modeling correctly predicted a distinct substructure well below the initial image resolution [see  \citet[Figs.~4~and~9]{hezaveh-etal2016} and \citet[Fig.~2]{spilker-etal2016}].  In this example, the prediction was reliable because the possible shapes of lensed sources are tightly constrained by the calculable effects of weak field lensing and the known properties of galaxies.  In the case of M87*, our knowledge of black hole lensing is comparably advanced---the photon ring properties are tightly constrained by established theory  \citep{gralla-holz-wald2019}---but our knowledge of the basic source structure is not.  Nevertheless, the demonstrated power of visibility-domain fitting---together with the fundamental importance of the photon ring---motivate a direct search in the M87* data.

We will refer to the main annular structure of the image as the ``annulus'' or ``primary ring'', and we refer to the thin ring component as the ``photon ring'' or ``secondary ring''. As we investigate the effect (or lack thereof) of the two additional model freedoms that have been introduced, we continue to focus on the fractional width of the primary ring. In general, the parameters of the primary ring are found to be independent of whether the photon ring is included. The diameter and orientation of the primary ring are very similar to what was found with the \textsf{xs-ring} model. The fractional width also remains at $f_w \leq 0.25$, and if anything is found to be even narrower than \textsf{xs-ring}, a surprising result given that the new model has more freedom to fit broader disks. This result affirms the narrowness of the image, favoring a more sharply peaked emitting region in the accreting plasma. Finally, we perform a test of the Bayesian evidence of models with and without a photon ring, and find that in some cases the analysis favors the model with a photon ring, while in others it favors the opposite.  Given these conflicting results, we conclude there is no evidence for the presence of a photon ring in the 2017 observations.

In Sec.~\ref{sec:model} we describe the construction of our new geometric model, and in Sec.~\ref{sec:results} we present results for its use on the 2017 data set.  In Sec.~\ref{comparison} we compare to other work.  We conclude and discuss some future directions in Sec.~\ref{conclusion}.  


\section{Refining the Model}\label{sec:model}

The interferometric data collected by the EHT comes in the form of \textit{visibilities}: complex-valued components of the Fourier transform of the underlying image brightness \citep{TMS}. When constructing a model for the image, the simplest route is to build it directly in the Fourier (aka visibility) domain, allowing for a computationally fast comparison between model and data. This is an advantage of the \textsf{xs-ring} model: its Fourier transform is analytic, so it can be expressed in the visibility domain from the beginning. To describe a disk with a slower radial falloff, we need to go beyond purely analytic models. 
Some components of the model can still be computed analytically -- in particular, the photon ring and any nuisance parameters -- but we will have to compute the visibilities of the disk numerically. So our strategy is semi-analytic: we start with a model of the primary ring in the image space, take a numerical Fourier transform, and then add the other components. We call this numerical disk model ``\textsf{ndisk}'', and when a photon ring is added on top, ``\textsf{ndisk+}''. We consider two choices for the radial brightness function: an exponential decay, and a power-law decay. The steepness of the falloff in either case is controlled by a single parameter, corresponding either to the exponential decay rate or the power law index.

\subsection{Construction of the NDISK+ Model}\label{construction}

To construct our model we begin with an exponentially decaying radial function, with the interior cut off by a step function at some radius $R_0$.  We add a dipolar brightness modulation to simulate the effects of Doppler beaming from an inclined disk of orbiting matter. The result is 
\begin{align}
    D_\textrm{exp}(x,y) = 
    \begin{cases} 
        \, 0, & r < R_0
        \\
        V_0 \, \mathrm{e}^{-m r / R_0} (1 + \beta x'/r) / I_0, & r \geq R_0 
    \end{cases},
    \label{Dexp}
\end{align}
where $x' = x \cos{\phi} + y \sin{\phi}$, $\phi$ is the direction of the gradient, $\beta$ controls the strength of the gradient, $V_0$ is the total flux density, and $I_0$ is a normalization constant given by
\begin{align*}
I_0 = 2 \pi R_0^2 \, \left( \frac{1+m}{m^2} \right) \, \mathrm{e}^{-m}.
\end{align*}
We compute the Fourier transform of the disk, $\tilde{D}(u,v)$, numerically (see App. \ref{app:resolution} for details), and then blur it with a Gaussian kernel of width $\sigma$. This blurring mainly serves to soften the \textit{inner} edge (previously a step function). Since the falloff far from the center is \textit{slower} than Gaussian, the rest of the disk is not affected by this blurring very much as long as $\sigma$ is not too large. 

Next we add a second component to the model: an infinitesimally thin ring of radius $R$ whose total flux density $V_0 A$ is a fraction $A$ of that of the main ring.  The Doppler boost from the source's orbital motion brightens the photon ring on the same side as the main ring (even though the rays are initially emitted backward), as seen in both analytic  \citep{gralla-lupsasca-marrone2020, chael-johnson-lupsasca2021} and general-relativistic magnetohydrodynamic (GRMHD) models \citep{johnson-etal2020}. We therefore consider a photon ring component with the same dipolar modulation,
\begin{align}
    Z(x,y) = \frac{A}{2\pi} \, V_0 \, \delta(r-R) \left( 1 + \beta \, \frac{x'}{r} \right).
\end{align}
The Fourier transform can be done analytically: 
\begin{align}
    \widetilde{Z}(u,v) = A \, V_0 \left( J_0(2 \pi R \rho) - \beta \, i \left(\frac{u'}{\rho}\right) J_1(2 \pi R \rho) \right).
\end{align}
Here $\rho = \sqrt{u^2+v^2}$ is the radial distance in Fourier space, $u' = u \cos{\phi} + v \sin{\phi}$ is a rotated coordinate, and $J_0$ and $J_1$ are Bessel functions. The final expression for the \textsf{ndisk+} model is therefore
\begin{align}
   \textsf{\textbf{ndisk+}}(u,v) = \mathrm{e}^{-2 \pi^2 \sigma^2 \rho^2} \widetilde{D}(u,v) + \tilde{Z}(u,v).
\end{align}

\begin{figure}
    \centering
    \includegraphics[width=\linewidth]{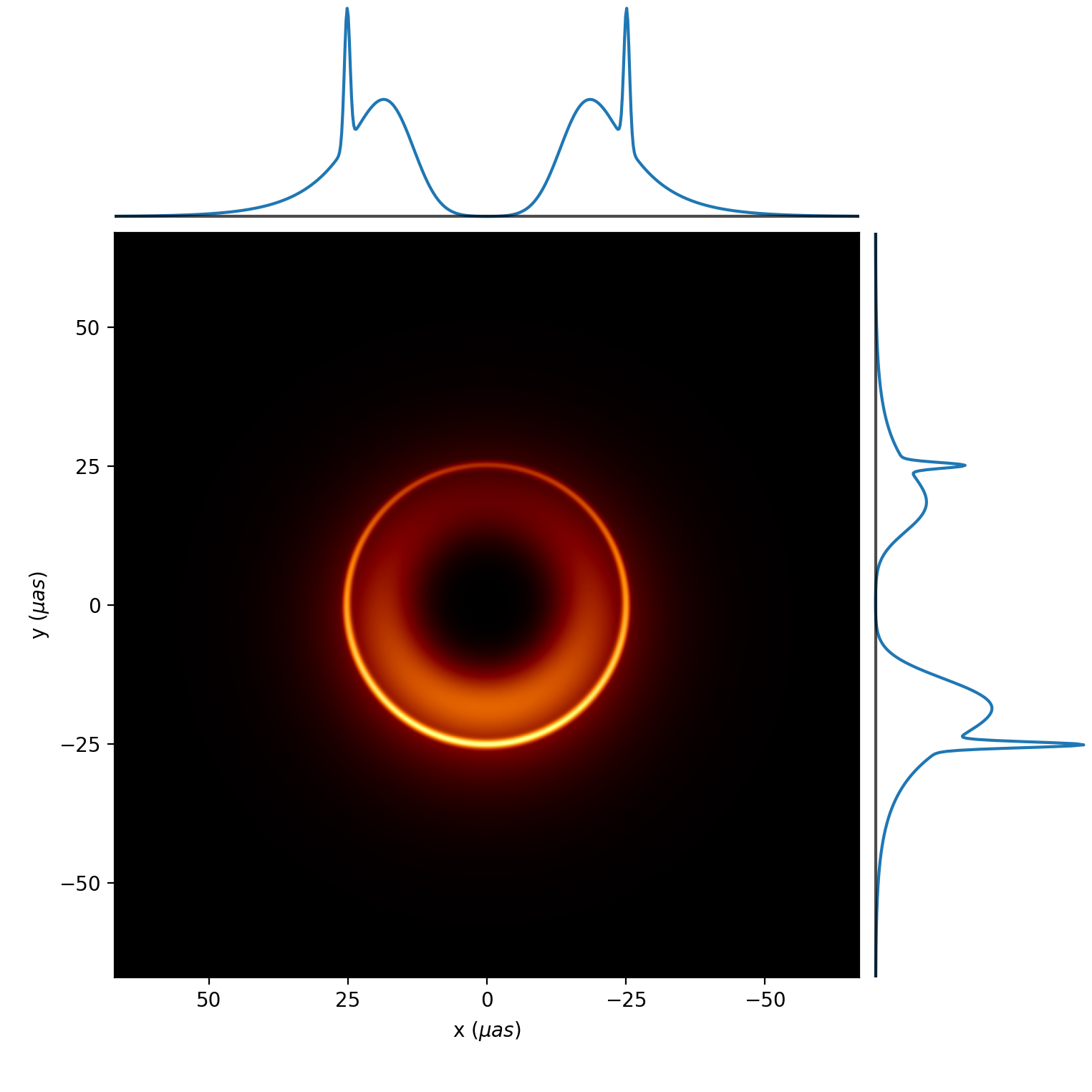}
    \caption{An example image generated from the \textsf{ndisk+} model with exponential decay. Parameters are: $\{R_0=15, m=2, \beta=0.4, \phi=3\pi/2, \sigma=4, A=0.1, R=25\}$. Cross-sections through $y=0$ and $x=0$ are shown above and to the right of the image respectively. The brightness gradient is clearly visible in the vertical cross-section. The photon ring – while truly infinitesimal in the model – has been blurred in this image with a Gaussian kernel of width $\sigma=0.5\mu$as for illustration, making its width and brightness consistent with expectations.
    }
    \label{fig:image}
\end{figure}

Fig.~\ref{fig:image} shows one realization of the \textsf{ndisk+} model. Taking a cross-section through the origin, the radial brightness profile has a sharp bump due to the photon ring, which lies just outside the peak brightness of the primary ring. This example looks similar to the profiles of GRMHD simulations in the MAD (magnetically arrested disk) configuration (\citetalias{EHT5}, \cite{johnson-etal2020}, \cite{chael-johnson-lupsasca2021}).  However, the model can also accommodate appearances similar to SANE (standard and normal evolution) disks, where the emission profile can be broader and the photon ring can be inside the central dark area \citepalias{EHT5}.  The \textsf{ndisk} model space also includes extreme cases such as filled disks, meaning it is broad enough to accommodate the expected observational appearance, while not ruling out surprises. 

The power-law version of the model is exactly the same, except for the original disk function $D(x,y)$. For a power-law falloff, we have instead

\begin{align}
    D_\textrm{pow}(x,y) = 
    \begin{cases} 
        0, & r < R_0 \\
        V_0 \, r^{-m} \, (1 + \beta x'/r) / I_0, & R_0 \leq r \leq R_\textrm{max} \\
        0, & r > R_\textrm{max}\label{Dpow}
    \end{cases}
\end{align}
Technically, the flux of this power-law disk does not converge for $m \leq 2$, so we cut off the disk at some very large radius $R_\textrm{max} = 1000 \mu$as (this somewhat arbitrary value was chosen to lie well outside the sampling window used to compute the Fourier transform -- see App.~\ref{app:resolution}). The normalization constant is now
\begin{align*}
I_0 = 2 \pi \frac{(R_\textrm{max})^{2-m} - (R_0)^{2-m}}{2-m}.
\end{align*}

Because we are fitting closure phases and closure amplitudes \citep{TMS, blackburn2020} (see Sec.~\ref{sec:results} below), the analysis is insensitive to the total flux in the image. Thus we can hold the flux of the disk fixed---we choose $V_0 = 1$Jy---and eliminate this degree of freedom from the model. (Closure quantities are also insensitive to translations of the image, so no model degree of freedom is needed for the location of the image center.) In total the model has 7 parameters: 5 for the primary ring ($R_0, m, \beta, \phi, \sigma$) and 2 for the photon ring ($A, R$). All model parameters are given uniform priors, which are listed in the following table: 

\vspace{5pt}
\begin{tabular}{ |c|l| } 
    \hline
    Parameter & Prior Range \\
    \hline
    $R_0$ & 1 -- 50 $\mu$as \\ 
    $m$ & 0 -- 8 \\ 
    $\beta$ & 0 -- 1 \\ 
    $\phi$ & 0 -- 2$\pi$ \\
    $\sigma$ & 1 -- 20 $\mu$as \\
    $A$ & 0 -- 0.25 \\
    $R$ & 1 -- 50 $\mu$as \\
    \hline
\end{tabular}
\vspace{5pt}

We cap the ratio of the photon ring to primary ring flux at $A \leq 25\%$. In models of geometrically thin disks in the Kerr spacetime, the de-magnification of the image that forms the photon ring is a factor of $10$--$20$ \citep{gralla-holz-wald2019, johnson-etal2020, gralla-lupsasca2020lensing}, corresponding to a ring with $\sim 5$--$10\%$ as much flux as the rest of the disk. With a thicker disk or more exotic emission mechanisms it may be possible for the flux ratio to be higher, but we are aware of no models in the literature where the ring flux exceeds $20\%$ \citep{gralla2021}. We enforce this upper limit for practical reasons, because if the photon ring  is allowed to take on too much flux we find there arises a swapping degeneracy between the photon ring and the primary ring. To avoid this complication we restrict the photon ring flux to physically realistic values. 

\subsection{Nuisance Parameters}

As the authors and the EHTC did before, we add nuisance parameters to the model in the form of elliptical Gaussians. For the time being, this is an unavoidable compromise in order to fit this dataset with such a simple model. Comparing results with different numbers of nuisance parameters tests the robustness of the claims, ensuring they are not simply an artifact of using an overly-simple model. Nuisance parameters are given the same priors as in the \textsf{xsring} model (see \citetalias{Lockhart-Gralla-2021}, Table A1). 

Lastly, we follow EHTC and add a very large-scale circular Gaussian, several orders of magnitude larger than the typical scale of the image \citepalias{EHT6}. This is intended to absorb unmodeled flux picked up by intra-site baselines, which enters into our analysis indirectly through closure amplitudes. The large-scale Gaussian is modeled with two free parameters as
\begin{align}
    \Tilde{G}(\rho) = V_G \, \mathrm{e}^{-2\pi^2 \sigma_G^2 \rho^2}.
\end{align}
This is the same as equation 45 of \citetalias{EHT6}. The flux $V_G$ is restricted between 0 and 10, while the width of the Gaussian $\sigma_G$ is allowed to vary from $10^{-2}$ to $10^1$ arc-seconds (not micro-arcseconds). We note that, while the large-scale Gaussian was omitted from \textsf{xs-ring} in \citetalias{Lockhart-Gralla-2021} after it was found to be irrelevant to the final results, here we have chosen to include it for completeness.

\subsection{Derived Parameters}

In our analysis we focus on five quantities $\{d, d_\textrm{ring}, \theta, f_w, m\}$ that are most relevant to the questions posed in this paper. They are: \\[4pt]
\renewcommand{\arraystretch}{1.2}
\begin{tabular}{ |r|l| } 
    $d :$ & diameter of the annulus \\ 
    $d_\textrm{ring} :$ & diameter of the photon ring \\ 
    $\theta :$ & position angle of brightest part of the annulus \\
    $f_w :$ & annulus fractional width \\ 
    $m :$ & falloff parameter \\[5pt]
\end{tabular}

Posterior distributions for these derived quantities are computed directly from the sampled model parameters. Three of them are trivial to calculate: the photon ring diameter is just twice its radius, $d_\textrm{ring} = 2R$; the position angle is the direction of the gradient measured East of North, which is just $\theta = 90^\circ-\phi$; and the falloff parameter $m$ is a model parameter already (it represents the falloff scale in the exponential model \eqref{Dexp}, and the power-law index in the power-law model \eqref{Dpow}). The other two, annulus diameter and fractional width, must be measured numerically in the image domain. To do this we rotate the image so that the gradient is in the horizontal direction, and then take a 1-dimensional slice in the vertical direction, perpendicular to the gradient. The diameter of the annulus $d$ is defined as the distance between the two peaks (note that this is \textit{not} simply $2R_0$, because the blurring of the inner edge of the annulus shifts the radius of peak brightness in a non-trivial way). The width of the annulus $w$ is defined as the full width at half maximum (FWHM) of each peak (the two peaks are identical because we sliced perpendicular to the gradient). The fractional width is then given by $f_w = w / d$.

\begin{figure*}
    \centering
    \includegraphics[width=0.95\textwidth]{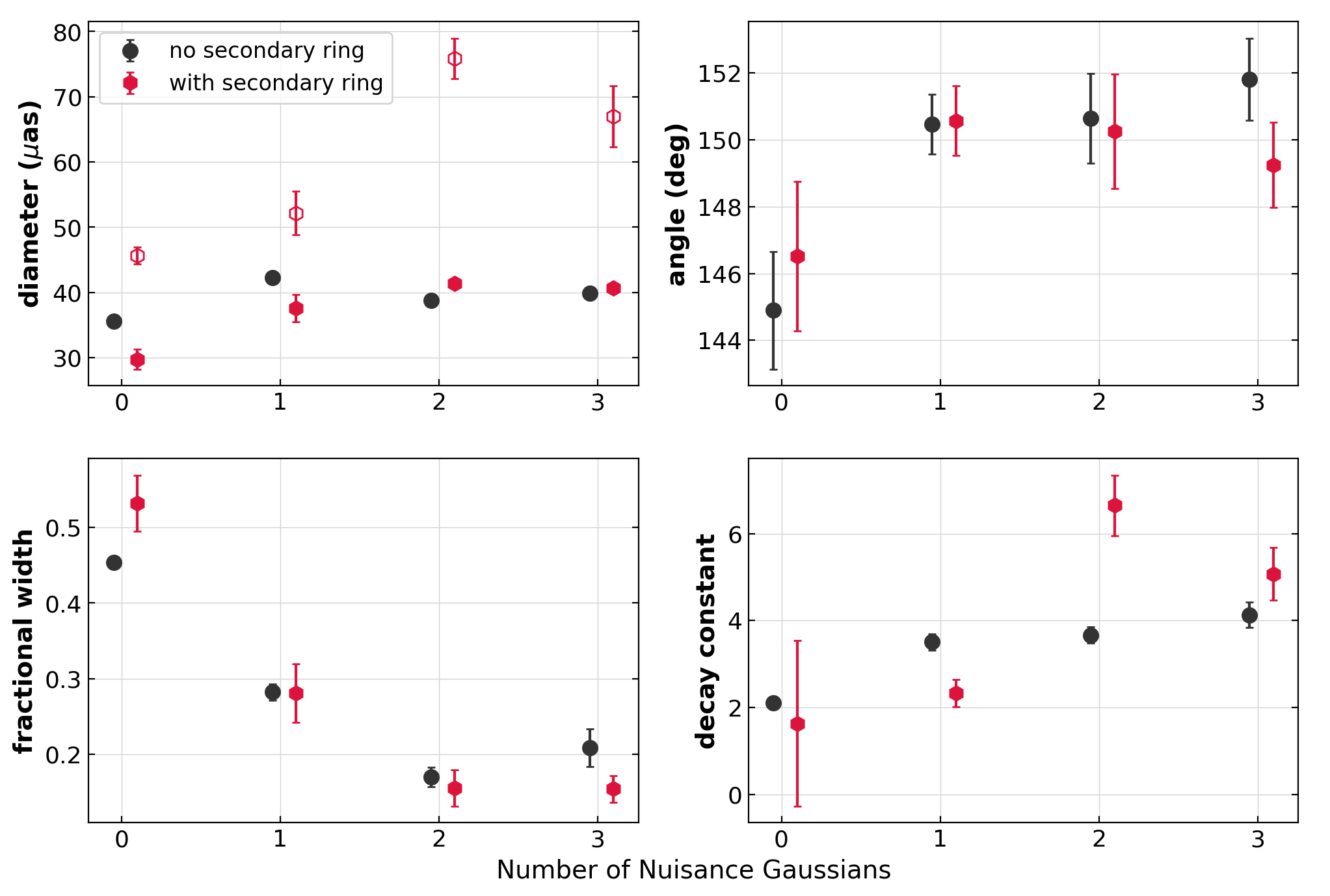}
    \caption{Posterior parameter ranges for the \textsf{ndisk} (no secondary ring) and \textsf{ndisk+} (with secondary ring) models with exponential falloff, using the April 06 hi-band dataset. We show the means and standard deviations of the diameter ($d$ and $d_\textrm{ring})$, position angle ($\theta$), fractional width ($f_w)$, and decay constant ($m$). For \textsf{ndisk+} runs, solid points on the diameter plot refer to the diameter of the main annulus $d$, while open points refer to the diameter of the secondary ring $d_\textrm{ring}$. Posteriors tend to stabilize after two nuisance Gaussians are added, while including a secondary photon ring has no qualitative effect on the posteriors of the primary ring.}
    \label{fig:apr06-exp}
\end{figure*}

\begin{figure*}
    \centering
    \includegraphics[width=0.33\textwidth]{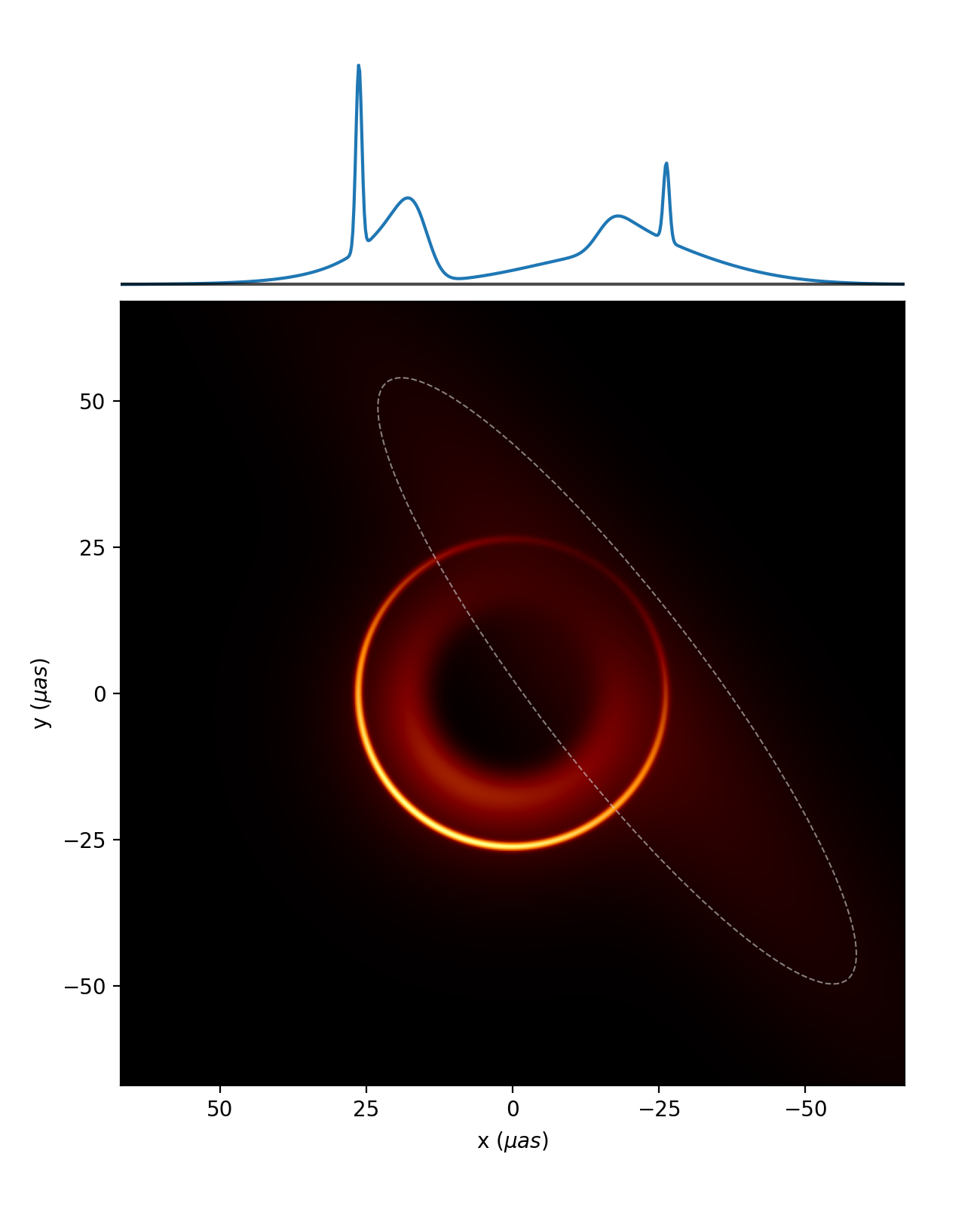}
    \includegraphics[width=0.33\textwidth]{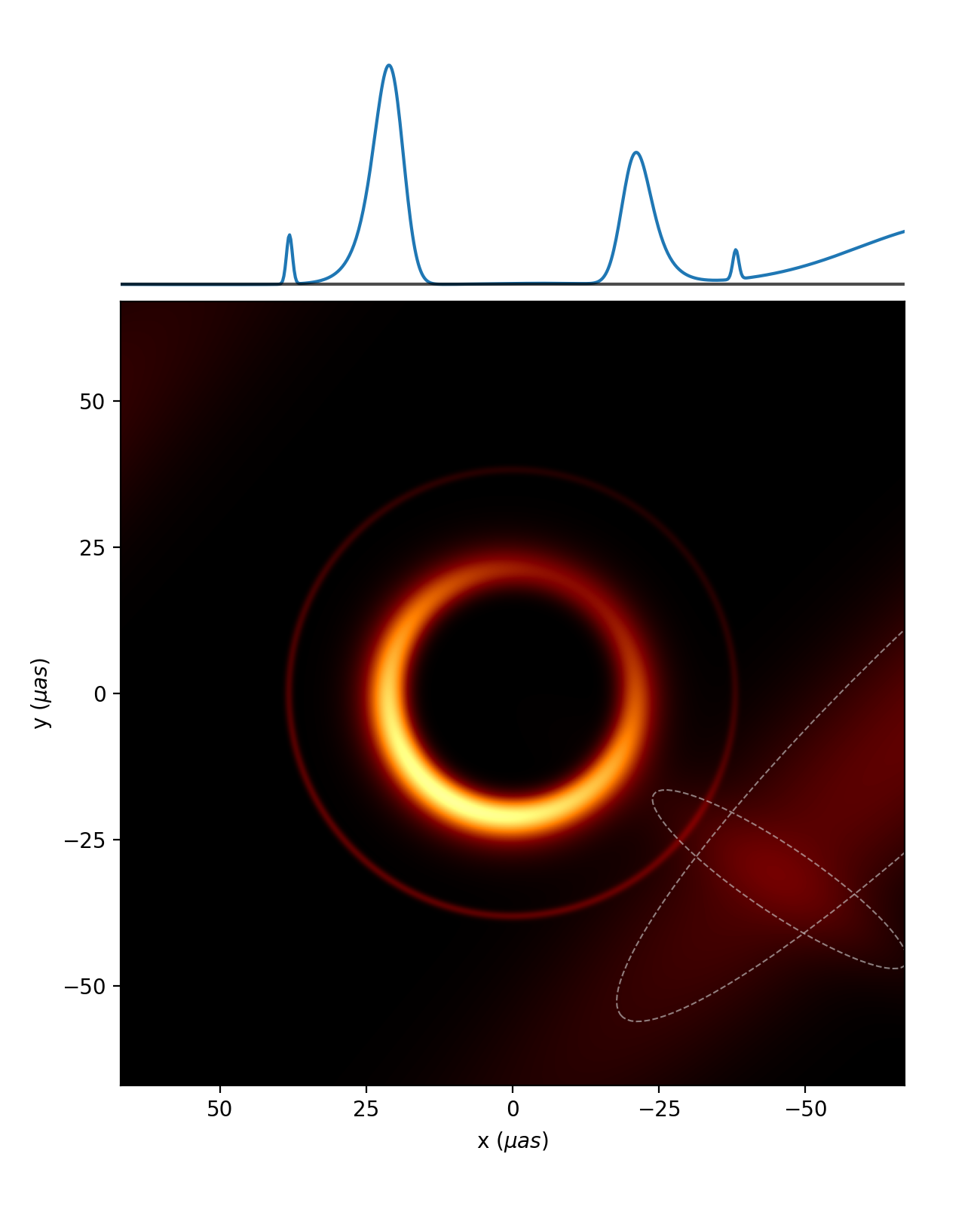}
    \includegraphics[width=0.33\textwidth]{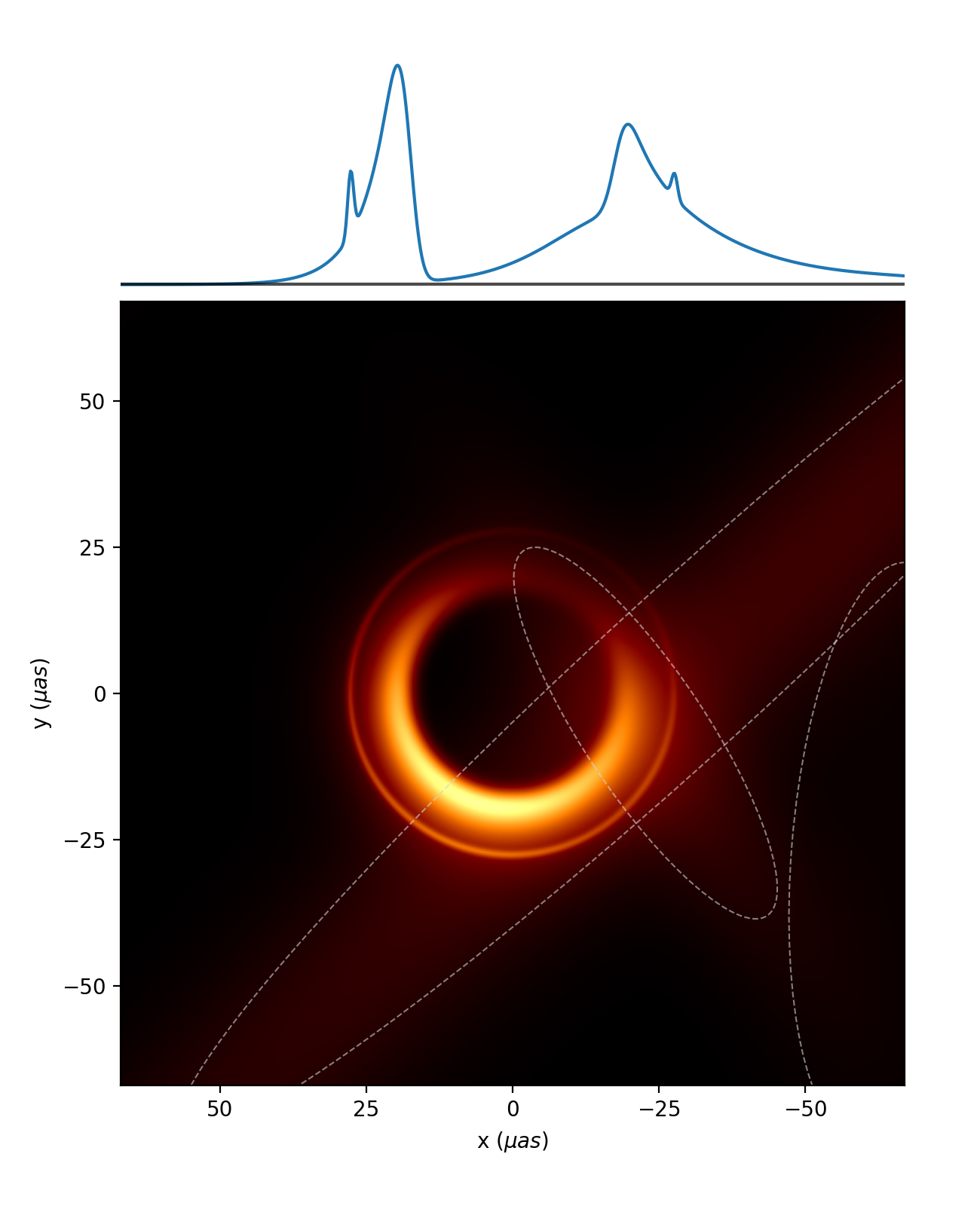}
    \caption{Best-fit images for runs that include a secondary ring (red points in Fig~\ref{fig:apr06-exp}). From left to right, the model has 1, 2, and 3 nuisance Gaussians. The FWHM of each Gaussian is shown as a dotted line, and horizontal cross-sections through the origin are shown above each image. The photon ring in each image has been blurred for illustration, as described in Fig.~\ref{fig:image}. The primary ring appearance is largely unchanged between 2 and 3 nuisance Gaussians, whereas the photon ring properties vary significantly.}
    \label{fig:best-fits}
\vspace{5pt}
\end{figure*}

\section{Results}\label{sec:results}

In this section we present the results of analyzing the \textsf{ndisk} and \textsf{ndisk+} models, with both exponential and power-law brightness falloff. The 2017 data includes four observation days in two frequency bands, for a total of 8 datasets. To compute model posteriors we use the dynamic nested sampling package \textsf{DYNESTY} \citep{dynesty2020}, which is well-suited for large parameter spaces and multi-modal distributions. Our approach to model fitting is described in detail in \citetalias{Lockhart-Gralla-2021}, Secs.~3 and 4. The closure phase and closure amplitude are used as the preferred data products to mitigate uncertainties coming from station calibration. The new models are fit using what is called in that paper the `fixed likelihood' function, the same as what was used in the EHTC analysis.

First we present posteriors for a single, representative dataset to show the effect of adding the photon ring component. Then we analyze the model across all datasets. We compare the Bayesian evidence for models with and without a photon ring, and discuss the fractional width and brightness falloff rate of the primary ring.

\subsection{A Representative Dataset}

\begin{figure}
    \centering
    \includegraphics[width=\linewidth]{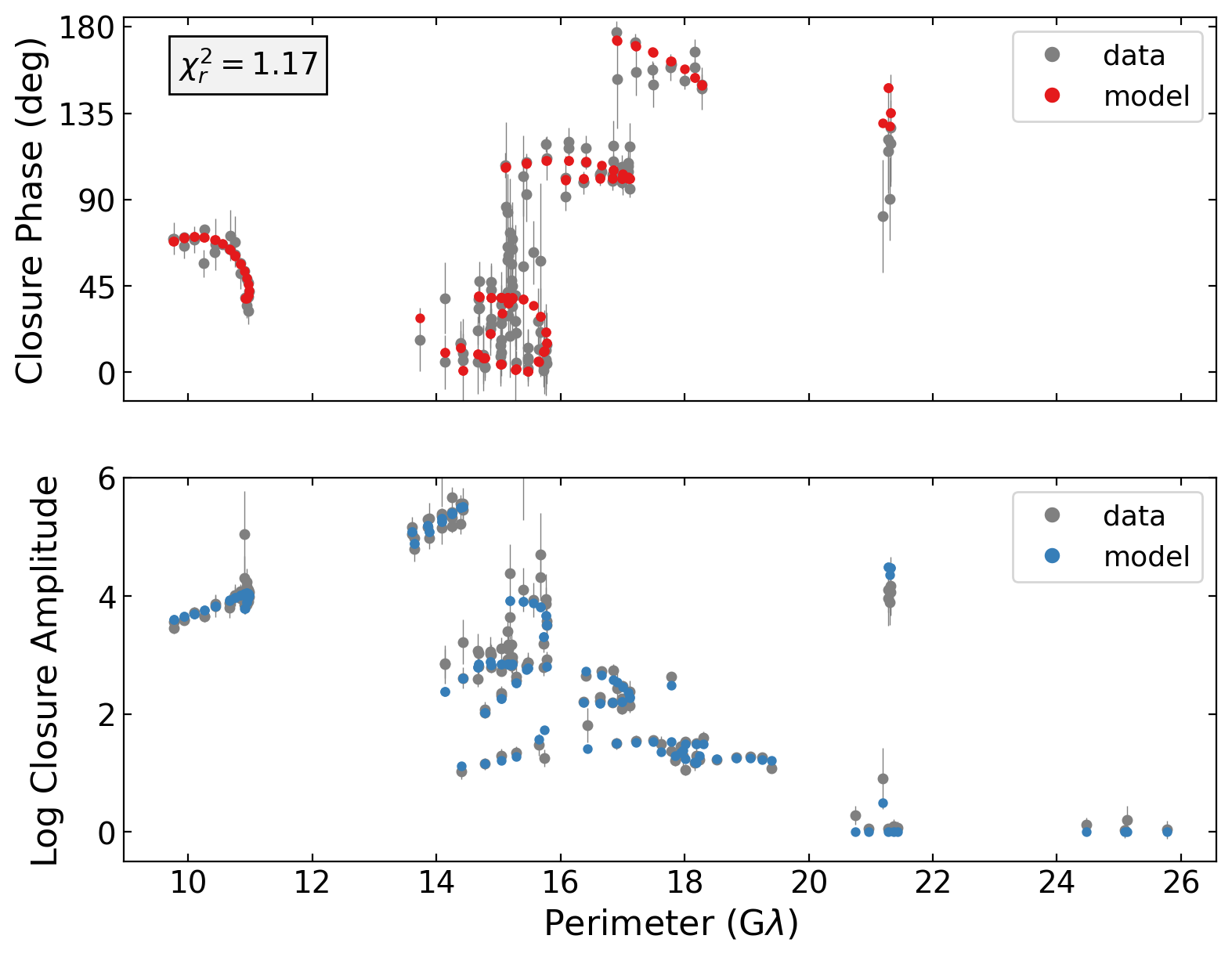}
    \caption{Best-fitting model and data points for an example run from Fig.~\ref{fig:apr06-exp}. This example has 2 nuisance Gaussians and no secondary ring. We plot the absolute values of the
    closure phase and log closure amplitudes as a function of the closure perimeter, defined as the sum of the lengths of the component baselines. This fit has a reduced chi-squared value of $\chi^2_r = 1.17$. (See \citetalias{Lockhart-Gralla-2021} for definitions of closure quantities and reduced chi-squared.)}
    \label{fig:closure-fits}
\end{figure}

We begin by examining in detail the April 06 hi-band dataset, which has the most data points and the highest overall signal-to-noise ratio.  We compare the \textsf{ndisk} and \textsf{ndisk+} models with exponential falloff, and four choices for the number of nuisance Gaussians included in the model ($0-3$). Posteriors for the key parameters of interest are represented in Figure~\ref{fig:apr06-exp}; images corresponding to the best-fit parameters from a few of these runs are shown in Figure~\ref{fig:best-fits}. Visual agreement with the data is shown for one example in Figure~\ref{fig:closure-fits}.\footnote{Defining what constitutes a `good fit' is subtle and somewhat fraught (see \citetalias[App.~C]{Lockhart-Gralla-2021}), but one commonly used metric is the reduced chi-squared. We find reduced chi-squared values near 1 when at least two Gaussians are included in the model, and the fit shown in Fig.~\ref{fig:closure-fits} has a chi-squared value that is comparable to those found with \textsf{xs-ring} by EHTC \citepalias{EHT6}.}

The primary ring parameters are similar whether the secondary ring is included or not, indicating that the secondary ring does not play a major role in determining the posteriors (Figure~\ref{fig:apr06-exp}). We note that the fractional width is especially robust, more so than the decay constant, whose variation is counteracted by other parameters of the model (the smoothing constant $\sigma$, for instance) to maintain a tight constraint on the fractional width. Posteriors tend to stabilize after 2 nuisance Gaussians are included.

On the other hand, the photon ring diameter and flux fraction vary greatly between runs, even after posteriors for the other main parameters have stabilized.  This behavior is similar to that of the unphysical nuisance Gaussians, suggesting that, for the 2017 EHT observations, the photon ring parameters are simply providing extra model freedom, and do not represent any real feature of the sky brightness.  But to properly assess whether the photon can be detected, we turn next to a different kind of test to specifically interrogate whether there is evidence for its presence in the data.

\subsection{Bayesian Evidence for a Photon Ring}\label{bayes-evidence}

To really test for the presence of a photon ring, we need to compare the efficacy of a model that includes it with one that does not. To do this we compute the marginal likelihood of each model, often called the `Bayesian evidence'. We now briefly review this approach, referring the reader to (e.g) \cite{trotta2008} for further details.

Let $p(A|B)$ denote the probability (density) of $A$ given $B$.  Consider a model $\mathcal{M}$ with parameters $\theta$ assigned prior probabilities $p(\theta|\mathcal{M})$.  For a set of data $d$, the model assigns a likelihood function $p(d|\mathcal{M},\theta)$ to each set of parameters $\theta$.  The model  \textit{evidence} $Z$ is defined to be the likelihood of the model itself, $p(d|\mathcal{M})$.  The evidence can be computed as an integral over the entire model space  $\Omega_\mathcal{M}$,
\begin{align}
    Z = p(d|\mathcal{M}) = \int_{\Omega_\mathcal{M}}  \, p(d|\theta,\mathcal{M}) p(\theta|\mathcal{M}) \, \textrm{d}\theta.
\end{align}
While the evidence does not provide absolute information about the quality of the model, relative information can be obtained if $\mathcal{M}$ exists in a larger space of models.  When only two models are considered, the evidence \textit{ratio} (sometimes called the ``Bayes factor'') is the factor by which the data changes the relative probability of the models.  If two models $\mathcal{M}_1$ and $\mathcal{M}_2$ are considered equally likely before the data are considered, the model $\mathcal{M}_2$ is considered $Z_2/Z_1$ times more likely after the data are taken into account.  For this reason, a ``large'' Bayes factor [e.g., greater than 150, so $\log|Z_2/Z_1|>5$ \citep{jeffreys1998theory}] is considered evidence that $\mathcal{M}_2$ is favored.

In the case where one model is a subset of the other, the Bayes factor is a way to quantify whether the additional parameters are pulling their weight. By integrating over the whole model space, a comparison of the evidence essentially reflects the \textit{average} improvement, so it is not overly influenced by the best fit.  In a sense, the Bayesian evidence is a means of quantifying the principle of Occam's razor: we want to balance the quality of the fit with the simplicity of the model. Using \textsf{DYNESTY} we can compute the Bayesian evidence for a model without a secondary ring, and a model with one.

\vspace{5pt}
\begin{tabular}{ |c|c|c| } 
    \hline
    Gaussians & Secondary Ring & Log Z \\
    \hline
    1 & no   & -18.0 $\pm$ 0.4 \\
    1 & yes & -17.6 $\pm$ 0.4 \\ 
    2 & no   & 81.6 $\pm$ 0.5 \\
    2 & yes & 85.7 $\pm$ 0.5 \\   
    3 & no   & 76.3 $\pm$ 0.5 \\
    3 & yes & 80.7 $\pm$ 0.5 \\ 
    \hline
    \\
\end{tabular}
\vspace{5pt}

\begin{figure}
    \centering
    \includegraphics[width=\linewidth]{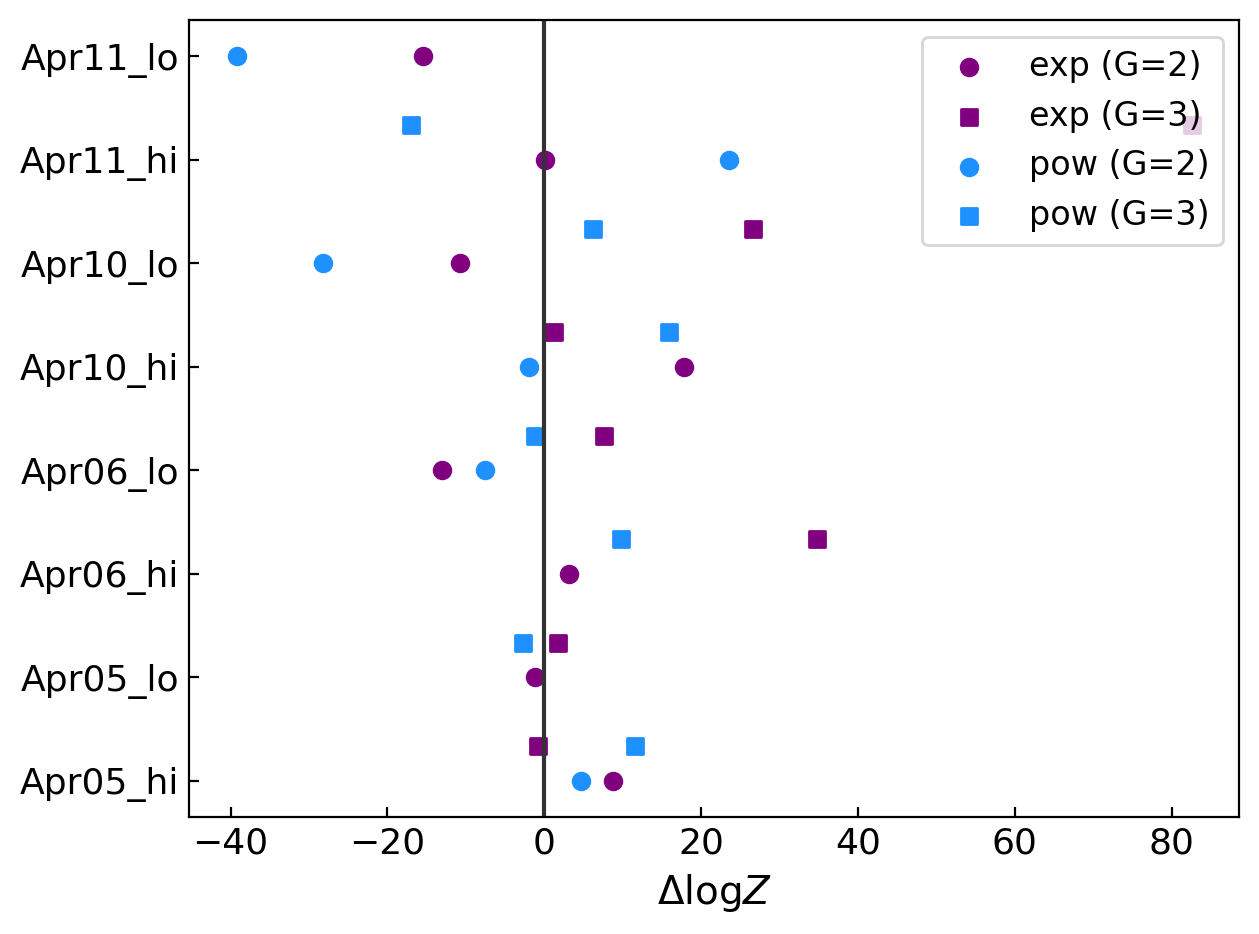}
    \caption{No evidence for a photon ring.  We show Bayes factors comparing models with and without a secondary ring. Two colors differentiate whether the falloff is exponential or power-law, and circles/squares indicate the number of nuisance Gaussians, either 2 or 3. $\Delta \log Z$, the difference in the logarithm of the evidence between two models, is the log of their Bayes factor. Positive values of $\Delta \log Z$ indicate evidence in favor of the secondary ring, while negative values mean the ring is disfavored. The results are split roughly evenly across the line $\Delta \log Z=0$, meaning there are as many cases for the photon ring as against.}
    \label{fig:Bayes}
\end{figure}

The table above shows the logarithm of the Bayesian evidence for different realizations of the model using the April 06-hi dataset. First we can see from these numbers that, whether a photon ring is included in the model or not, the evidence increases dramatically when a second nuisance Gaussian is added, while the difference is marginal upon adding a third. This pattern is repeated in the other datasets as well. This tracks with the observation that the posteriors converge after two Gaussians (Fig.~\ref{fig:apr06-exp}), and suggests that 2-3 is the most appropriate number to include to improve the flexibility of the model without over-fitting. Some indication of this was given in \citetalias{Lockhart-Gralla-2021} (Fig. C1) comparing the likelihoods of the best-fit models in each case, but this is a more convincing result. 

To examine the evidence for the photon ring, we restrict to just the 2- and 3-Gaussian cases, and compute the Bayes factor between the models with and without the photon ring  for exponential and power-law falloff models across all datasets. The results are shown in Fig.~\ref{fig:Bayes}. Overall, evidence for a photon ring is unconvincing; there are as many instances where the photon ring is favored as where it is disfavored. Ultimately this is an inconclusive result: we can neither confirm nor rule out the presence of a photon ring around M87*. Prospects for a future detection are discussed in Sec.~\ref{comparison}.

\begin{figure}
    \centering
    \includegraphics[width=\linewidth]{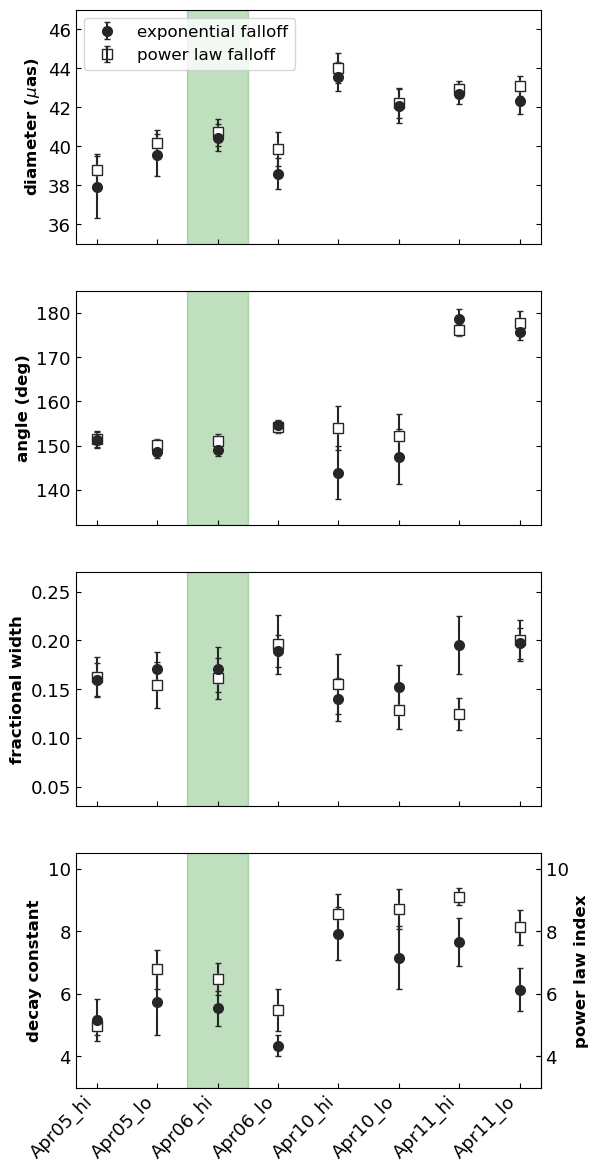}
    \caption{Posterior ranges for the numerical disk model across all datasets, with either exponential or power-law falloff. Each point represents an average of 4 individual runs: including either 2 or 3 nuisance Gaussians, with or without a secondary ring. Points represent the average of the means of each run, and error bars represent the average of the standard deviations. The diameter and orientation angle are consistent with previous modeling efforts, while the fractional width remains as low or lower than past results. In the bottom plot, the decay constant for the exponential disk model (black) is plotted alongside the power law index for the power law disk model (white). The April 06-hi dataset featured in Figures~\ref{fig:apr06-exp}, \ref{fig:best-fits}, and \ref{fig:closure-fits} is highlighted in green.}
    \label{fig:results}
\end{figure}
\subsection{Full Results for the Primary Ring}

We will now examine what the full set of data can tell us about the primary ring. Figure~\ref{fig:results} summarizes the results of our analysis across all 8 datasets. We find that the diameter and brightness orientation angle of the disk are very similar to the results for the \textsf{xs-ring} model reported in \citetalias{Lockhart-Gralla-2021}, as long as at least one nuisance Gaussian is included in the latter. This supports the previously reported value of $d \sim 40 \mu$as, with a brighter side towards the Southern part of the sky. Once again, the photon ring degree of freedom is largely irrelevant to constraining the properties of the main ring.

Most interesting, and most relevant to the topic of this series, is that the fractional width remains low---mean value less than or equal to $0.2$---even for this model. This is important because it need not have been this way; this model has the freedom to choose a much broader disk decaying at a leisurely rate, a regime that was inaccessible to \textsf{xs-ring}. Yet despite the added freedom in the falloff rate (and a secondary ring component), the fractional widths are even a bit smaller than before.

The decay constant for both the exponential and power-law models shows significant variation across observation days, with mean values ranging from $4$ to $9$.  However, this entire range represents significantly faster falloff than seen in models of M87*.  For example, Fig.~7 of  \citet{chael-johnson-lupsasca2021} shows a roughly exponential falloff in brightness for a radiative GRMHD model.  Measuring the slope by eye and picking $R_0 \approx 15 \mu$as to compare with our model \eqref{Dexp}, we see that the effective falloff parameter $m$ is no larger than $2$.  As a second example, consider the semi-analytical models presented recently in \citet{Vincent2022}, which have a roughly exponential emission profile $\sim e^{-3 r/r_H}$, where $r$ is the Boyer-Lindquist coordinate and $r_H$ is the horizon radius.  This translates to a similar decay rate $m \approx 3$ in the image domain (see App.~D of \citetalias{Lockhart-Gralla-2021} for some heuristics).  In other words, the observed brightness falloff is consistently more rapid than seen in models of the source.

\section{Other methods for photon ring detection}\label{comparison}

We now compare our work with other proposed methods for detecting the photon ring.  The most closely related proposal is the geometric/pixel hybrid model of \cite{broderick2020}, which was very recently applied to the 2017 observations \citep{broderick2022}.\footnote{The manuscript appeared while this paper was under review, with the review process nearing completion.}  We will refer to this as the Broderick method.  While the authors of \citet{broderick2022} regard their analysis as a detection of the photon ring, we instead view their results as further confirmation that there is no evidence for the photon ring in the 2017 observations.  We now compare the methods in detail and explain our point of view.

The Broderick method is similar in spirit to ours, in that one considers an image model with and without an additional photon ring component and seeks evidence that inclusion of the photon ring improves the fit quality.  However, there are a number of technical differences between the methods, with varying degrees of importance.  

First, the data products and samplers differ between the approaches.  While no detailed comparison has been attempted, we do not expect either approach to be significantly better than the other.  Second, the model for the main image component differs: the Broderick method considers a sparse grid of pixels using interpolation to make a smooth image, while we consider an annulus with additional Gaussian components.  Both models offer some flexibility with respect to the main emission using a similar total number of parameters, and it is difficult to say whether one is superior to the other.

A more important difference is the criteria used to assess whether the inclusion of a photon ring improves the fit.  Whereas we use the Bayes factor, the Broderick method instead relies on the Bayesian information criterion (BIC).  Although the BIC approximates the Bayes factor under some assumptions \citep{raftery1995bayesian}, it is not clear whether these assumptions are satisfied for the model under consideration.  The Broderick method also considers the Akaike information criterion, which does not have a Bayesian interpretation.  We would also note that, even admitting the validity of these information criteria, the evidence in favor of the ring component is rather marginal, occurring in only three out of the four observation days \citep{broderick2022}.

The most important difference between the methods is the choice of priors for the photon ring component parameters.  Our method restricts the photon ring flux density to be less than $25\%$ of the main ring flux density.  This removes degeneracy with the main emission and ensures that the photon ring is subdominant, as it must be on general grounds \citep{gralla-holz-wald2019}.  By contrast, the Broderick method allows unphysically large values of the photon ring flux, and in fact \textit{finds} unphysical values of $\sim$54--64\% for the 2017 observations \citep{broderick2022}.  This suggests that the ``photon ring'' component is simply helping to model the main ring-like emission by providing a significant pure-ring contribution.\footnote{The prescribed narrowness of the photon ring component is not an obstacle to this interpretation, since the baselines involved are too short to resolve such narrow structures.}

The Broderick approach was able to roughly reproduce photon ring diameters from mock data generated from a handful of GRMHD simulations \citep{broderick2020}.  However, in all of these simulations the photon ring is directly on top of the main emission, so there is no way to check whether the ``photon ring'' model  component is in fact sensitive to the photon ring, as opposed to simply representing a portion of the main ring-like emission.  In fact, the method finds the same unphysically large flux fractions discussed in the previous paragraph (i.e., it does \textit{not} recover the correct photon ring flux density of the  underlying model), supporting the interpretation that the supposed photon ring component simply reflects a portion of the main emission.

Based on the above reasoning, we interpret the analysis of \citet{broderick2022} as showing that a grid+ring model is marginally preferred to a pure grid model for the 2017 observations.  This is not surprising since the M87* appearance is known to feature a ring.  This analysis says nothing about the presence of a photon ring.  

We are thus in the unfortunate situation that two independent methods have failed to uncover evidence of a photon ring in the 2017 observations.  These methods use rather different models for representing the main emission, and yet support the same conclusion.  While we leave open the possibility that a vastly improved model for the main emission might help reveal a photon ring in the 2017 observations, it seems likely that detection of the photon ring will instead have to await significant improvements in data quality.

An alternative approach to measuring the photon ring was proposed by \citet{johnson-etal2020}.  The idea is to directly measure the visibility-domain signature of the ring on (sufficiently long) baselines where it dominates the signal.  This method is remarkably insensitive to astronomical uncertainties and can even provide the \textit{precise shape} of the photon ring \citep{gralla2020, gralla-lupsasca2020observable}, allowing precision tests of general relativity \citep{gralla-lupsasca-marrone2020, paugnat2022}. However, reaching the required baselines would almost certainly require a space mission.

\section{Conclusion} \label{conclusion}

We have constructed a new geometric model for the M87* near-horizon image, which improves over previous models by incorporating a parameterized radial brightness profile and a component describing the photon ring predicted by general relativity. We find that this model agrees with previous studies on the diameter and position angle of the main annular structure, and still corroborates the relatively narrow fractional width of $f_w \leq 0.25$ found with the \textsf{xs-ring} model. Evidence for the presence of a photon ring coming from the Bayes factor of our models is inconclusive, and the photon ring degrees of freedom have no effect on the main conclusions. 

Let us therefore return to the title of this series of papers: How narrow is the M87* ring?  The EHTC analysis bounded the fractional width to be less than one-half, with the geometric models favoring narrow rings, and the imaging models favoring thicker rings.  One possible explanation for the narrow rings in the geometric models is that these models lack the freedom to accommodate a gradual falloff rate, and compromise by presenting a narrow ring at the peak brightness of the true ring.  Similarly, one could imagine that the presence of a photon ring on top of the main emission confuses the model into thinking the main ring is narrower than it is.  The analysis of this paper rules out these explanations: the ring remains narrow when the models are given the relevant additional freedom.

Adopting the viewpoint that modeling is more reliable than imaging for determining fine features in the sky appearance, the evidence is accumulating in favor of a narrow ring in M87*.   A thin ring is rather dissimilar from theoretical expectations  \citepalias{Lockhart-Gralla-2021}, and could point to the need for a revision in our understanding of the accretion flow.  However, it should be borne in mind that an alternative likelihood function does favor somewhat thicker rings \citepalias{Lockhart-Gralla-2021}. In this series of papers we have elected to change one component of modeling at a time (the likelihood function in the first paper; the geometric model here), in order to keep the analysis manageable and test the importance of the different modeling choices.  Ultimately, it may be necessary to wait for improved observations, for which differences in modeling assumptions may have less impact on the final conclusions.

Another important limitation of the geometric models is their use of unphysical nuisance parameters.  The elliptical Gaussians are motivated by convenience, rather than physical arguments, and it is hard to say whether they are capturing unmodeled features of the image, or simply represent systematic uncertainty.  A more physically motivated choice of additional model freedom would allow increased confidence in the conclusions of the analysis.  We plan to revisit this question in future papers in the series. 

In the coming years, more radio telescopes will be added to the EHT array, higher-frequency observations will be conducted, and sensitivity will be improved \citep{ngEHT}.  We hope that the techniques introduced in this paper will help pin down the width of the M87* ring and perhaps even enable a detection of the photon ring component.


\section*{Acknowledgements}

This work was supported in part by NSF grant PHY1752809 to the University of Arizona.


\section*{Data Availability}

The data analyzed in this study are described in \citetalias{EHT3} and are available as \href{https://doi.org/10.25739/g85n-f134}{DOI:10.25739/g85n-f134}. See the section on data availability in \citetalias{Lockhart-Gralla-2021} for a note on conventions.


\bibliographystyle{mnras}
\bibliography{EHTdataII} 


\appendix

\section{Numerical Resolution}\label{app:resolution}

To compute the Fourier transform of the model we take a Discrete Fourier Transform (DFT) via the Fast Fourier Transform algorithm. An important consideration is whether the DFT is computed at sufficient resolution. In fact, there are two parameters that control the fidelity of the transformation: the size of the pixels, and the size of the sampling window. Since the flux in our models technically extends out to infinity, we need to check that both the pixel size and the sampling window are sufficient to prevent any unwanted artefacts. To do this we tested our DFT algorithm on functions with known analytic Fourier transforms, and checked that with these settings the parameters we infer from our model are converged.

\begin{figure}
    \centering
    \includegraphics[width=\linewidth]{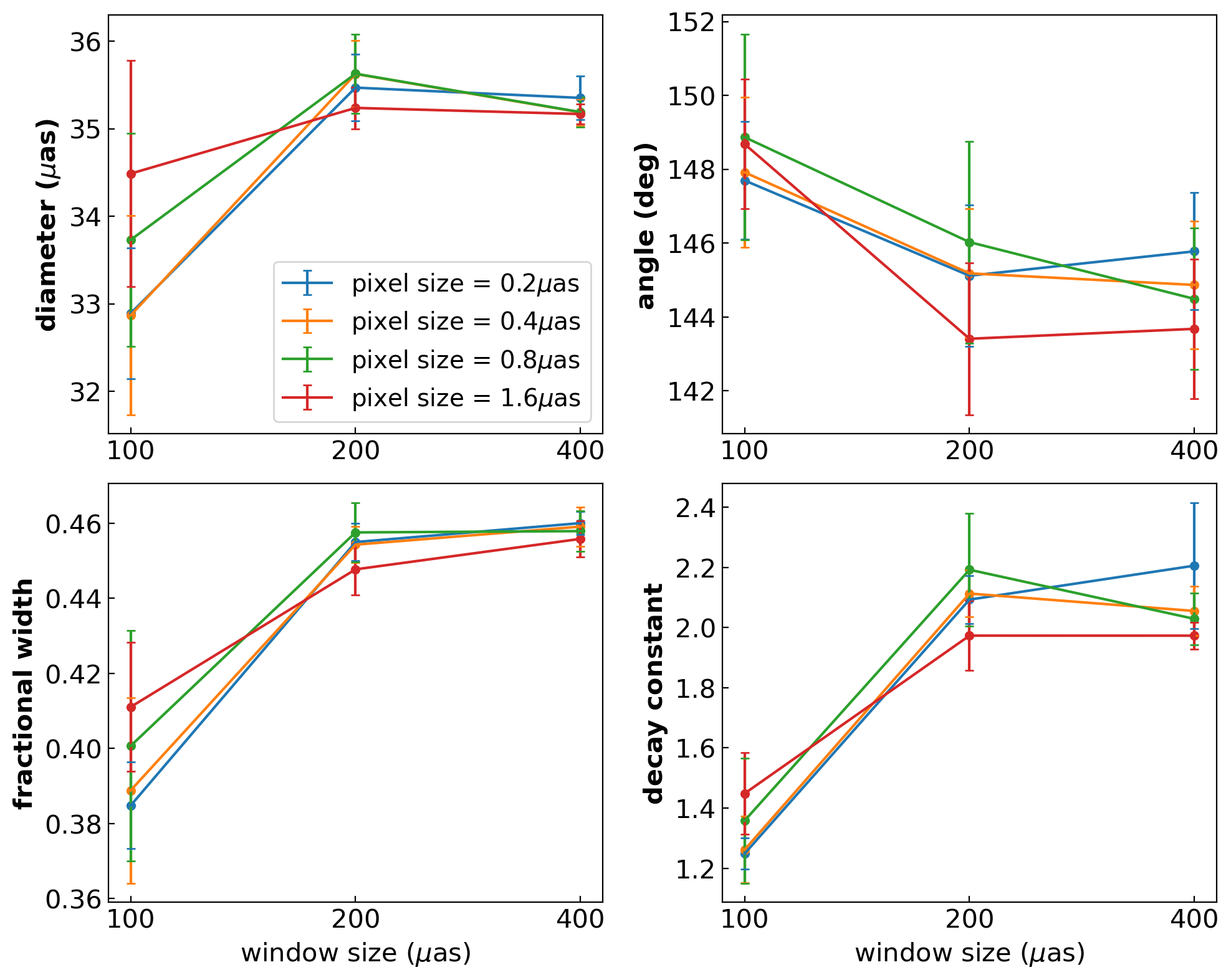}
    \caption{Mean and standard deviation of key model parameters found using different pixel sizes and sampling window sizes, for the Apr 06 hi-band dataset using the \textsf{ndisk} model. Values converge to within the error bars for a pixel size of $0.4\mu$as (orange) and a sampling window of $200\mu$as.}
    \label{fig:resolution}
\end{figure}

Fig~\ref{fig:resolution} shows posteriors for the \textsf{ndisk} model with exponential falloff as a function of resolution for the Apr 06 hi-band dataset. Note that this is only the \textit{numerical} part of the model - the primary ring - and does not include the photon ring or nuisance parameters, which are both computed analytically (hence why some of the values differ from those shown in the main results of the paper). The resolution used in our analysis was a window size of $200 \mu$as and a pixel size of $\approx 0.4\mu$as ($200 \mu$as / 512 pixels). From the figure above it can be seen that neither increasing the pixel density (cutting the pixel size in half from $0.4$ to $0.2 \mu$as), nor enlarging the sampling window (doubling from $200$ to $400 \mu$as) has much effect on the posteriors. Therefore we conclude that this resolution is sufficient.


\bsp	
\label{lastpage}
\end{document}